\documentclass{PoS}

\usepackage{amsmath}
\usepackage{epsfig}
\usepackage{empheq}
\usepackage{xcolor}
\usepackage{graphicx}
\def\bea#1\eea{\begin{align}#1\end{align}}

\newcommand{\nnu}{\nonumber\\}
\newcommand{\bef}{\begin{figure}[htb]\centering}
\newcommand{\eef}{\end{figure}}
\newcommand{\hx}{\hat{x}}
\newcommand{\hz}{\hat{z}}
\newcommand{\pht}{p_{h\perp}}
\newcommand{\xb}{x_B}

\newcommand{\nslash}{\kern 0.2 em n\kern -0.50em /}
\newcommand{\kslash}{\kern 0.2 em k\kern -0.45em /}
\newcommand{\pslash}{\kern 0.2 em p\kern -0.50em /}
\newcommand{\Sslash}{\kern 0.2 em S\kern -0.50em /}
\newcommand{\Pslash}{\kern 0.2 em P\kern -0.50em /}
\newcommand{\Dslash}{\kern 0.2 em D\kern -0.65em /\kern 0.15em}

\title{Three-gluon correlator contribution to Sivers asymmetry in SIDIS }

\ShortTitle{Three-gluon correlator contribution to Sivers asymmetry in SIDIS}

\author{\speaker{Ling-Yun Dai}\\
        Center For Exploration of Energy and Matter, Indiana University, Bloomington, IN 47408, USA\\
        Physics Department, Indiana University, Bloomington, IN 47405, USA\\
        E-mail: \email{lingyunphysics@gmail.com}}

\abstract{We study the Sivers asymmetry in semi-inclusive deep inelastic scattering: the three-gluon correlation function contribution. At the cross section level, we establish the matching between the twist-3 collinear factorization approach and the transverse momentum dependent factorization formalism at the moderate transverse momentum region. The so-called coefficient functions when one expands the quark Sivers function in terms of three-gluon correlation functions are derived, which are essential components in the usual transverse momentum dependent evolution formalism. Finally we calculate the transverse-momentum-weighted spin-dependent differential cross section at the next-to-leading order, from which we identify the off-diagonal/three-gluon correlation contribution to the QCD collinear evolution of the twist-3 Qiu-Sterman function. }

\FullConference{QCD Evolution 2015 -QCDEV2015-\\
		26-30 May 2015\\
		Jefferson Lab (JLAB), Newport News Virginia, USA}

\begin{document}

\section{Introduction}
Transverse-spin asymmetries have become important tools to probe the novel nucleon structure, in particular the parton's transverse motion~\cite{Boer:2011fh,Accardi:2012qut,Aschenauer:2013woa,Aschenauer:2015eha}. On the experiment side, $e+p$ \cite{Qian:2011py,Zhao:2014qvx,Airapetian:2009ae,Adolph:2014zba}, $p+p$ \cite{Adare:2010bd,Adamczyk:2012kn} and $e^+e^-$ annihilation \cite{Seidl:2008xc,TheBABAR:2013yha} processes have been performed to study these asymmetries. On the theory side, two approaches have been developed:
the so-called transverse momentum dependent (TMD) factorization \cite{Ji:2004wu,Ji:2004xq,Collins:2011zzd} for low transverse momentum of the observed hadron, and the twist-3 collinear factorization approach \cite{Efremov:1981sh,Efremov:1984ip,Qiu:1991pp} for high transverse momentum. The relation  between these two mechanisms (so-called ``matching'') has been discussed for Drell-Yan production in \cite{Ji:2006ub} as well as for semi-inclusive deep inelastic scattering (SIDIS) process in \cite{Bacchetta:2008xw}, both of which involve the quark-gluon correlation function or the so-called Qiu-Sterman function. In Sec.~II below we will discuss the matching for photon-gluon fusion channel of SIDIS at cross section level, which involves the three-gluon correlation function.

SIDIS is an efficient experimental tool to study the single transverse-spin asymmetries (SSAs).
In this process, SSAs generated by Sivers function are rather interesting and extensively studied experimentally by HERMES \cite{Airapetian:2009ae}, COMPASS \cite{Adolph:2014zba,Alekseev:2008aa,Adolph:2013stb} and JLab \cite{Qian:2011py}. Sivers function is very intriguing due to the theoretical prediction that it will change sign in Drell-Yan with respect to SIDIS process~\cite{Brodsky:2002cx,Collins:2002kn,Boer:2003cm,Kang:2009bp}. A number of experiments including COMPASS, RHIC, and Fermilab are planned to test this prediction experimentally. Knowledge of evolution of the Sivers function~\cite{Sivers:1989cc,Kang:2011mr,Aybat:2011ge,Echevarria:2012pw,Sun:2013hua,Echevarria:2014xaa} with hard scale is very important for accurate phenomenological application and eventually for our better knowledge of these functions. To have the full knowledge of TMD evolution of the Sivers function, one needs to know the so-called coefficient functions. In Sec.~III, we derive a set of such coefficient functions when one expands the quark Sivers function in terms of the three-gluon correlation functions. Finally in Sec.~IV we study the next-to-leading order (NLO) perturbative QCD corrections to the transverse momentum-weighted spin-dependent cross section. By analyzing the collinear divergence structure, we identify the off-diagonal evolution kernel of the Qiu-Sterman function.

\section{Matching between twist-3 collinear factorization and TMD formalism}
We study the three-gluon correlation contribution to the Sivers asymmetry in SIDIS process: $e(\ell)+p(p, s_\perp)\to e(\ell')+h(p_h)+X$. With slightly different normalization from Ref.~\cite{Beppu:2010qn}, the three-gluon correlation function is defined as follows:
\bea
M^{\alpha\beta\gamma}_{F,abc}(x_1,x_2) &= g_s \int\frac{dy_1^-dy_2^-}{2\pi}  e^{i x_1 p^+ y_1^-}  e^{i (x_2-x_1) p^+ y_2^-} \frac{1}{p^+} \langle PS| F_b^{\beta +} (0)F_c^{\gamma +} (y_1^-)F_a^{\alpha +} (y_2^-)  |PS\rangle   \nonumber \\
&=  \frac{N_c}{(N_c^2-1)(N_c^2-4)} d^{abc} O^{\alpha\beta\gamma}(x_1,x_2)  - \frac{i}{N_c(N_c^2-1)} f^{abc} N^{\alpha\beta\gamma}(x_1,x_2)\; ,
\label{eq:MF}
\eea
where $O^{\alpha\beta\gamma}(x_1,x_2)$ and  $F^{\alpha\beta\gamma}(x_1,x_2)$ correspond to symmetric and anti-symmetric combinations of gluon field-strength tensors. The transverse spin dependent differential cross section of Sivers effect, or the so-called $\sin(\phi_h - \phi_s)$ module, could be written as
 \bea
\frac{d \Delta\sigma}{dx_B dy dz_hd^2 {\bf p_{h\perp}}}= \frac{\alpha_{em}^2 y}{32 \pi^3 Q^4 z_h} L^{\mu\nu} W_{\mu\nu} (p,q, p_h),
\label{eq:crs}
\eea
where $L^{\mu\nu} = 2(\ell^\mu \ell'^\nu + \ell^\nu \ell'^\mu) - Q^2 g^{\mu\nu}$ is the leptonic tensor, while $W^{\mu\nu}$ is the hadronic tensor, including the partonic tensor $w^{\mu\nu}$ and the usual
fragmentation function $D_{h/q}(z)$. The generic Feyman diagram to calculate the partonic hard-part function is sketched in Fig.~\ref{fig:amplitude}. In the following calculation, we only consider the so-called metric contribution~\cite{Daleo:2004pn}, i.e., we contract our hadronic tensor with ($-g^{\mu\nu}$).

%%%%%%%%%%%%%%%%%%%%%%%%%%%%%%%%%%%
\bef
\psfig{file=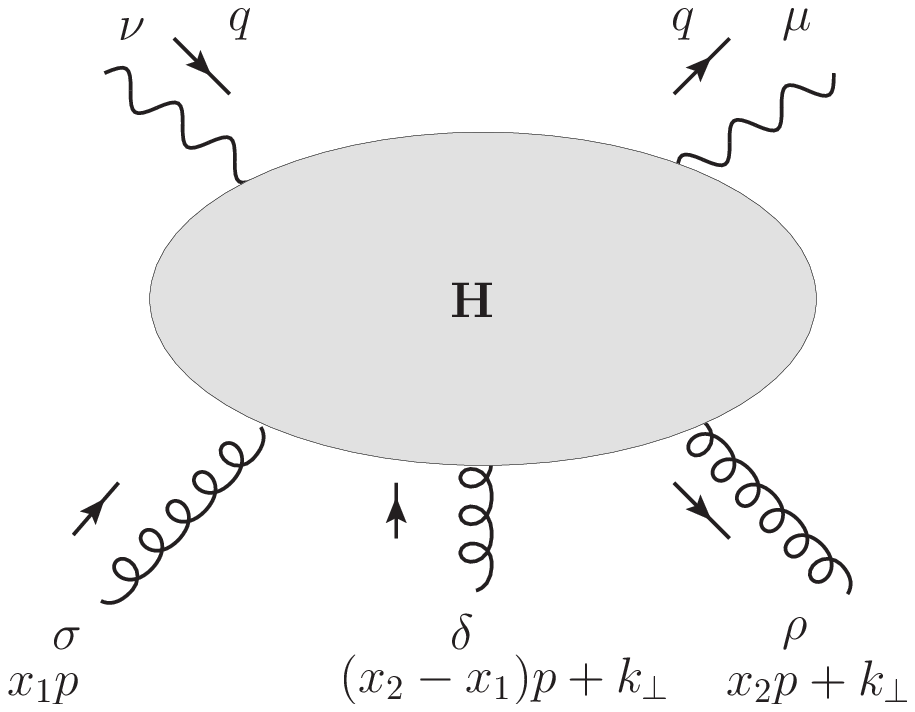, scale=0.45}
\caption{Generic diagram that is used to calculate the hard-part function $H_{\rho\delta\sigma}^{abc}$.}
\label{fig:amplitude}
\eef
%%%%%%%%%%%%%%%%%%%%%%%%%%%%%%%%%%%
One can compute the contribution of the three-gluon correlation function to the spin-dependent cross section up to $\mathcal{O}(\alpha_{\rm em}\alpha_s)$,
which has already been studied in \cite{Beppu:2010qn} for heavy meson production. Here what we are interested is to study its matching to the TMD factorization formalism. Because of different goals, we will simply derive the results for {\it light} hadron production, i.e., neglect the mass of the hadrons. We found that in the low $\pht \ll Q$ limit, the spin-dependent cross section can be written as
\bea
\left.\frac{d\Delta\sigma}{d\xb dy dz_h d^2 {\bf p_{h\perp}}}\right|_{\pht \ll Q} =& - z_h \sigma_0 \left(\epsilon^{\alpha\beta} s_\perp^\alpha \pht^\beta \right) \frac{1}{\left(\pht^2\right)^2} \sum_q e_q^2
\frac{\alpha_s}{2\pi^2} \int \frac{dz}{z} D_{h/q}(z) \delta(1-\hz)
\nnu
&\times \int \frac{dx}{x^2} P_{q\leftarrow g}(\hx) \left( \frac{1}{2}\right) \left[ O(x,x) + O(x,0) + N(x,x) - N(x,0)\right],
\label{eq:twist-3}
\eea
where $P_{q\leftarrow g}(\hx) = T_R \left[\hx^2 + (1-\hx)^2\right]$ is the usual gluon-to-quark splitting kernel,
with $T_R = 1/2$ the color factor. On the other hand, the TMD factorization formalism \cite{Ji:2004wu,Ji:2004xq,Collins:2011zzd} for SIDIS process reads
\bea
\frac{d\Delta\sigma}{d\xb dy dz_h d^2{\bf p_{h\perp}}} =& \sigma_0 \sum_q e_q^2 \int d^2{\bf k_\perp} d^2{\bf p_\perp} d^2{\bf \lambda_\perp} \delta^2\left(z_h {\bf k_\perp}+{\bf p_\perp}+{\bf\lambda_\perp}-{\bf p_{h\perp}} \right)
\nnu
&\times
\frac{\epsilon^{\alpha\beta} s_\perp^\alpha k_\perp^\beta}{M}f_{1T}^{\perp q}(x_B, k_\perp^2) D_{h/q}(z_h, p_\perp^2) S(\lambda_\perp) H(Q^2),
\label{eq:TMD-fac}
\eea
where $f_{1T}^{\perp q}(x_B, k_\perp^2)$ is the quark Sivers function, $D_{h/q}(z_h, p_\perp^2)$ is the TMD fragmentation function,
$S(\lambda_\perp)$ and $H(Q^2)$ denotes the soft and hard factors, respectively. To study the connection between the TMD approach and the collinear twist-3 formalism, as given in Eqs.~(\ref{eq:TMD-fac}) and (\ref{eq:twist-3}), we expand the quark Sivers function $f_{1T}^{\perp q}(x_B, k_\perp^2)$ in terms of the three-gluon correlation functions at $k_\perp \gg \Lambda_{\rm QCD}$,
\bea
\frac{1}{M}f_{1T}^{\perp q}(x_B, k_\perp^2) = -\frac{\alpha_s}{2\pi^2} \frac{1}{\left(k_\perp^2\right)^2} \int_{x_B}^1 \frac{dx}{x^2}  P_{q\leftarrow g}(\hat x) \left(\frac{1}{2}\right)\left[O(x,x) + O(x,0) + N(x,x) - N(x,0)\right].
\label{eq:sivers-expand}
\eea
For other factors,  we let $k_\perp$ to be of the order of  $p_{h\perp}$  and the others ($\lambda_\perp$ and $p_\perp$) much smaller \cite{Ji:2006br}. In this case, we neglect $\lambda_\perp$ and $p_\perp$ in the delta function:
\bea
k_\perp\sim \pht: \qquad \delta^2\left(z_h {\bf k_\perp}+{\bf p_\perp}+{\bf\lambda_\perp}-{\bf p_{h\perp}}\right) \rightarrow \delta^2\left(z_h {\bf k_\perp} - {\bf p_{h\perp}} \right).
\eea
At the same time using:
\bea
\int d^2{\bf \lambda_\perp} S(\lambda_\perp)=1,
\qquad
\int d^2{\bf p_\perp} D_{h/q}(z_h, p_\perp^2) = D_{h/q}(z_h),
\eea
and $H(Q^2)=1$, finally we find that from Eq.~(\ref{eq:TMD-fac}) we arrive at the same result of Eq.~(\ref{eq:twist-3}). We thus demonstrates the matching between collinear twist-3 factorization formalism and TMD factorization formalism for three-gluon correlation functions at moderate transverse momentum, $\Lambda_{\rm QCD} \ll \pht \ll Q$.

\section{Coefficient functions in TMD evolution formalism}
We will now derive the coefficient functions associated with the three-gluon correlation functions, which are essential ingredients in the usual TMD evolution formalism. Since TMD evolution formalism is often derived in the Fourier transformed coordinate $b$-space (conjugate to $k_\perp$ in momentum space), such coefficient functions usually relate the quark Sivers function in the $b$-space to the corresponding collinear functions.

To start, we recall that the quark Sivers function in the  the coordinate $b$-space is defined as follows~\cite{Kang:2011mr,Echevarria:2014xaa}:
\bea
f_{1T}^{\perp q(\alpha)}(x_B, b) = \frac{1}{M} \int d^{2-2\epsilon} {\bf k}_\perp e^{-i {\bf k}_\perp\cdot {\bf b}} k_\perp^\alpha f_{1T}^{\perp q}(x_B, k_\perp^2).
\eea
Here the expression is given in $n=4-2\epsilon$ dimensions, in which our calculations will be performed to regularize the potential divergences as we will show below. In other words, in the perturbative region $1/b \gg \Lambda_{\rm QCD}$, one can expand the quark Sivers function as a product of the coefficient functions $C_{q\leftarrow i}(\hat x_1, \hat x_2)$ and the corresponding collinear function $f^{(3)}(x_1, x_2)$, {\it i.e.}, the twist-3 Qiu-Sterman function $T_{q,F}(x_1, x_2)$ as well as the three-gluon correlation functions $O(x_1, x_2)$ and $N(x_1, x_2)$:
\bea
f_{1T}^{\perp q(\alpha)}(x_B, b) =  \left(\frac{ib^\alpha}{2}\right) C_{q\leftarrow i}(\hat x_1, \hat x_2)\otimes f_{i}^{(3)}(x_1, x_2).
\eea
where $\otimes$ represents the convolution over the momentum fractions.
At leading order (LO), the result is given by
\bea
f_{1T}^{\perp q(\alpha)}(x_B, b) = \left(\frac{ib^\alpha}{2}\right) T_{q,F}(x_B, x_B),
\label{eq:bare}
\eea
where the quark Sivers function is for SIDIS process~\cite{Echevarria:2014xaa}. If it is for Drell-Yan process, there will be an extra sign on the right hand side of the equation~\cite{Kang:2011mr}. At the next-to-leading order, we have
\bea
f_{1T}^{\perp q(\alpha)}(x_B, b) = & \left(\frac{ib^\alpha}{2}\right)
\Bigg\{
\frac{\alpha_s}{2\pi} \left(-\frac{1}{\hat \epsilon}\right) \int \frac{dx}{x^2} P_{q\leftarrow g}(\hat x) \left(\frac{1}{2}\right)\left[O(x,x) + O(x,0) + N(x,x) - N(x,0)\right]
\nonumber\\
&+ \int_{x_B}^1\frac{dx}{x^2} \Big[C_{q\leftarrow g, 1}(\hat x) \big(O(x, x)+N(x,x)\big)+ C_{q\leftarrow g, 2}(\hat x) \big(O(x,0) - N(x,0)\big) \Big]\Bigg\},
\eea
where $1/\hat \epsilon=1/\epsilon-\gamma_E+\ln 4\pi$. It is obvious that the first term (the divergent term) is simply the $\mathcal O(\alpha_s)$ correction to the Qiu-Sterman function $T_{q,F}(x,x)$~\cite{Collins:2011zzd,Bacchetta:2013pqa}, and it should be recombined with Eq.~(\ref{eq:bare}) to define the renormalized Qiu-Sterman function. The rest finite terms correspond to the desired coefficient functions, with the expressions given by
\bea
C_{q\leftarrow g,1}(\hx) &= \frac{\alpha_s}{4\pi} \left[P_{q\leftarrow g}(\hx) \ln\left(\frac{c^2}{b^2\mu^2}\right)+\hx(1-\hx)\right], \nonumber
\\
C_{q\leftarrow g,2}(\hx) &= \frac{\alpha_s}{4\pi} \left[P_{q\leftarrow g}(\hx) \ln\left(\frac{c^2}{b^2\mu^2}\right)-\frac{1}{2}\left(1-6\hx+6\hx^2\right)\right],
\label{eq:C}
\eea
where $c=2e^{-\gamma_E}$. We thus have derived the coefficient functions $C_{q\leftarrow g}$ of off-diagonal part: the contribution of three-gluon correlation functions.
Such coefficient functions will be exactly {\it the same} even if one uses the new properly defined TMDs in \cite{Collins:2011zzd} and/or \cite{Echevarria:2012js,Echevarria:2012pw}, because there is no contribution from soft factor subtraction at order ${\mathcal O}(\alpha_s)$ \cite{Collins:2011zzd, Bacchetta:2013pqa} for the off-diagonal part.

\section{Transverse momentum weighted spin-dependent cross section}
In this section we study the NLO transverse momentum-weighted transverse spin-dependent cross section, which is defined as~\cite{Kang:2012ns}:
\bea
\frac{d\langle \pht \Delta\sigma\rangle}{dx_B dy dz_h}
\equiv
\int d^2{\bf p_{h\perp}} \epsilon^{\alpha\beta} s_{\perp}^\alpha \pht^{\beta} \frac{d \Delta\sigma}{dx_B dy dz_h d^2{\bf p_{h\perp}}}.
\label{pht-weight}
\eea
The $\pht$-weighted cross section contains collinear divergence, to regularize such a divergence, we again present all the calculations in $n=4-2\epsilon$ dimensions. We then perform the usual $\epsilon$-expansion to isolate the divergent and finite contributions. Collecting these terms, and performing integration by parts to convert all the derivative terms $df^{(3)}(x_1,x_2)/dx$ to non-derivative terms $f^{(3)}(x_1,x_2)$, we end up with the following expression:
\bea
\frac{d\langle\pht\Delta\sigma\rangle}{d\xb dy dz_h} =& -\frac{z_h\sigma_0}{2} \sum_q e_q^2\int \frac{dz}{z}
D_{h/q}(z) \delta(1-\hz) \left(-\frac{1}{\hat \epsilon} + \ln\left(\frac{Q^2}{\mu^2}\right)\right)
\nnu
&\times
\frac{\alpha_s}{2\pi} \int \frac{dx}{x^2} P_{q\leftarrow g}(\hx) \left(\frac{1}{2}\right) \left[ O(x,x) + O(x,0) + N(x,x) - N(x,0)\right] + \cdots,
\label{eq:divergent}
\eea
where ``$\cdots$'' represents the finite NLO corrections and are suppressed here. The details are given in our paper \cite{Dai:2014ala}. Comparing Eq.~\eqref{eq:divergent} with the LO result given by~\cite{Kang:2012ns}
\bea
\frac{d\langle \pht \Delta\sigma\rangle}{dx_B dy dz_h} = -\frac{z_h\sigma_0}{2} \sum_q e_q^2
\int \frac{dx}{x} \frac{dz}{z} T_{q,F}(x,x) D_{h/q}(z) \delta(1-\hat x)\delta(1-\hat z),
\label{lo-res}
\eea
we notice that the divergent part is simply the NLO collinear QCD correction to the LO bare Qiu-Sterman function $T_{q,F}^{(0)}(x_B, x_B)$.
It should be absorbed into the definition of the renormalized $T_{q,F}(x_B, x_B)$.
After $\overline{\rm MS}$ regularization we obtain the evolution equation for the Qiu-Sterman function of the off-diagonal (three-gluon correlation function) contribution:
\bea
\frac{\partial}{\partial \ln\mu_f^2} T_{q,F}(x_B, x_B, \mu_f^2) &= \frac{\alpha_s}{2\pi}\int_{x_B}^1 \frac{dx}{x^2}  P_{q\leftarrow g}(\hat x) \left(\frac{1}{2}\right)\left[O(x,x,\mu_f^2) + O(x,0,\mu_f^2) \right.\nonumber\\
&\hskip 1.6in\left. + N(x,x,\mu_f^2) - N(x,0,\mu_f^2)\right].
\eea
This result agrees with the earlier findings~\cite{Kang:2008ey,Braun:2009mi,Ma:2012xn}.
After such a subtraction, we finally obtain the NLO corrections (finite parts) of the three-gluon correlation functions to the $\pht$-weighted transverse spin-dependent differential cross section. The complete result can be found in \cite{Dai:2014ala}. The result follows the standard form in the usual collinear factorization.

\section{Summary}
In this talk we report our recent work on the three-gluon correlation function contribution to the Sivers asymmetry in SIDIS~\cite{Dai:2014ala}. At cross section level, we demonstrate the matching between twist-3 collinear factorization formalism and TMD factorization at moderate hadron transverse momentum, $\Lambda_{\rm QCD}\ll \pht \ll Q$. We also derive the ${\mathcal O}(\alpha_s)$ expansion of the quark Sivers function in terms of the three-gluon correlation functions, the so-called off-diagonal piece. From it we derive the coefficient functions used in the usual TMD evolution formalism. We further calculate the NLO perturbative QCD corrections to the transverse-momentum-weighted spin-dependent differential cross section, from which we identify the QCD collinear evolution of the twist-3 Qiu-Sterman function: the off-diagonal contribution from three-gluon correlation functions.

\section*{Acknowledgments}
\noindent I thank Z. Kang, A. Prokudin and I. Vitev for collaborations and thank Y. Koike for helpful discussions. This work is supported by Indiana University College of Arts and Sciences and the U.S. Department of Energy, Office of Science, Office of Nuclear Physics under contract DE-AC05-06OR23177.

\end{document}